\documentclass[aps,prd,twocolumn,showpacs,notitlepage,eqsecnum,
superscriptaddress,nofootinbib]{revtex4-2}

\usepackage[centertags]{amsmath} \usepackage{amssymb} \usepackage{latexsym}
\usepackage{enumerate} \usepackage{graphicx} \usepackage{mathrsfs}
\usepackage[colorlinks]{hyperref} \usepackage{stmaryrd} \usepackage{import}
\usepackage{tensor} \usepackage[usenames,dvipsnames]{xcolor} \usepackage{bm}
\usepackage{multirow} \usepackage{breakurl} \usepackage{float}
\usepackage{verbatim} \usepackage{mathrsfs}
\usepackage{booktabs}
\usepackage{orcidlink}

\allowdisplaybreaks[1]

\definecolor{CiteColor}{rgb}{0,0.5,0} \hypersetup{citecolor=CiteColor}
\definecolor{RefColor}{rgb}{0.55,0,0} \hypersetup{linkcolor=RefColor}
\definecolor{darkgreen}{rgb}{0.2,0.7,0.2}
\definecolor{red}{rgb}{1,0,0}

\begin{document}

\title{Schwarzschild perturbations in Lorenz gauge via elliptic differential equations }

\newcommand{\geneseo}{\affiliation{Department of Physics and Astronomy, State University of New York at Geneseo, New York 14454, USA}}
\newcommand{\ucd}{\affiliation{ School of Mathematics and Statistics, University College Dublin, Belfield, Dublin 4, Ireland}}
\newcommand{\rpi}{\affiliation{Department of Physics, Applied Physics, and Astronomy, Rensselaer Polytechnic Institute, Troy, New York 12180, USA}}

\author{Thomas Osburn\,\orcidlink{0000-0003-2747-3994}} \geneseo \ucd
\author{Barry Wardell\,\orcidlink{0000-0001-6176-9006}} \ucd
\author{Erin Battaglia\,\orcidlink{0009-0008-5692-9747}} \geneseo \rpi

\date{\today}

\begin{abstract}

Accurate predictions of gravitational wave signals from asymmetric compact binaries are accessible through black hole perturbation theory and self-force calculations. Faithful waveform models will require contributions from first- and second-order terms in the small mass-ratio expansion. The problem of second-order Kerr perturbations is exacerbated by non-separability of the metric perturbation equations and non-linear mode coupling, which motivates this $m$-mode approach. This work moves towards the eventual goal of second-order Kerr perturbations by calculating first-order Schwarzschild metric perturbations via $m$-modes in the frequency domain for the first time. We solve the Lorenz gauge field equations as a system of coupled elliptic partial differential equations that govern each $m$-mode. Our Mathematica code implements a second-order finite difference representation of the field equations, which we solve as a sparse linear algebra problem. Regularization near the small body is achieved through the effective source method, and our presentation introduces a new puncture expansion of the singular field for a point mass in Kerr spacetime. Issues related to problematic near-horizon behavior are explored and then mitigated by applying sophisticated near-horizon boundary conditions. Our results illustrate the features of each component and $m$-mode of the metric perturbation, and we are able to calculate gravitational wave energy fluxes with sufficient accuracy to enable future second-order self-force calculations.

\end{abstract}

\maketitle

\section{Introduction}

The bandwidth of detectable gravitational wave (GW) frequencies will 
expand with the launch of the Laser Interferometer Space Antenna (LISA)~\cite{LISA,LISA:2024hlh,LISAWaveformWG:2023arg}, a 
space-based GW detector designed to probe signals between 0.1~mHz to 0.1~Hz. This
newly-accessible frequency range sits between the audible-band frequencies currently probed by the ground-based
GW detectors of the LIGO-Virgo-KAGRA detector network~\cite{Abbott_2019,Abbott_2021,Abbott_2022,Abac_2026,LIGO5} and the ultra-low nanoHertz frequencies achievable through pulsar timing array techniques~\cite{Agazie_2023,EPTA_2023,Reardon_2023,Xu_2023}. Furthermore, the deployment of LISA will likely coincide with that of the Einstein Telescope~\cite{ET:2025xjr}, Cosmic Explorer~\cite{Evans:2021gyd}, Taiji~\cite{Taiji}, and/or TianQin~\cite{Luo_2016}, which will assemble a powerful array of third-generation GW detectors. This fertile environment will provide ample opportunities to observe new GW sources with stronger signal-to-noise ratios and novel features.

Asymmetric compact binaries in the extreme mass-ratio inspiral (EMRI) and intermediate mass-ratio inspiral (IMRI) configurations are valuable and interesting~\cite{EMRI} examples of these new GW sources. These asymmetric mass-ratio systems involve two compact objects with masses $\mu$ and $M$ in the range $\sim 1-10^8 \; M_\odot$ such that $q \equiv M/\mu \gg 1$. It is valuable to accurately model these systems through the loudest strong-field phase of the GW signal, which is achievable through numerical relativity or black hole perturbation theory calculations. Although numerical relativity calculations are being pushed to higher $q$ ranges through clever advancements~\cite{Fernando_2019,Lousto_2023,Wittek_2025,mahapatra_2026}, it is well known that catastrophic scale disparities cause numerical relativity to be impractical above a certain intermediate $q$ ceiling. In contrast, the black hole perturbation theory and gravitational self-force methods~\cite{Mino_1997,Quinn_1997,Poisson:2011nh,Barack:2018yvs,Pound_2021} applied in this work are well-suited to describe these asymmetric systems through an expansion in powers of $\epsilon \equiv 1/q$ (although, a $q$ floor is encountered at some intermediate value instead). Accurate LISA waveform models will require both first-order ($\sim \epsilon$) and second-order ($\sim \epsilon^2$)~\cite{Pound_2005,Burke_2024} gravitational perturbations. Recent work in the Lorenz gauge based on separation of variables (the $lm$-mode approach~\cite{Barack_2005,Barack_2007,Barack_2010,Akcay_2011,Akcay_2013,Osburn_2014,Hopper_2015,Wardell_2015,Dolan_2022,Dolan_2024,Wardell_2025,Thornburg_2026}) led
to successfully calculating second-order Schwarzschild metric perturbations for the first time \cite{Pound_2020,Wardell_2023,Albertini_2022,vandemeent_2023}. To improve astrophysical realism, it will be necessary to account for the spin of the primary by solving the significantly more difficult problem of second-order Kerr gravitational perturbations. 

There are a number of reasons why the second-order Kerr problem is more challenging than the second-order Schwarzschild problem. Many of these challenges are related to Kerr's reduced symmetry and are mitigated by abandoning full separation of variables and retaining both the $r$ and $\theta$ dependence in each mode coefficient, which is known as the $m$-mode approach. A through explanation of the advantages provided by $m$-modes for second-order Kerr calculations are presented in~\cite{Vu_2026}, and we briefly summarize them here. First, there is no known equivalent of a tensor spherical harmonic basis in which the Kerr metric perturbation equations directly separate (although there are indirect workarounds being investigated~\cite{Green_2020,Toomani_2021,Dolan_2022,Dolan_2024,Wardell_2025,Mei_2026,mei2026separatinglinearizedeinsteinequations}). Second, with $lm$-modes the mode-coupling problem of the second-order source becomes exacerbated~\cite{Spiers_2024} compared to $m$-modes (and is worse for Kerr compared to Schwarzschild).

In contrast, with $m$-modes the Kerr second-order source construction procedure is relatively efficient, and work is underway to build $m$-mode sources for second-order Kerr metric perturbations. Originally, the $m$-mode approach was developed for time-domain black hole perturbation theory and self-force calculations~\cite{Barack_2007a,Barack_2007b,Dolan_2011a,Dolan_2011b,Dolan_2013,Thornburg_2017,Thornburg_2020}. Here we adopt the frequency-domain because it is well adapted to the multiscale expansion involved in second-order self-force calculations and because time-domain evolutions in the Lorenz gauge have encountered unchecked growth in time~\cite{Barack_2007,Barack_2010,Dolan_2013} (although recent $lm$-mode advancements to avoid this issue are likely generalizable to $m$-modes~\cite{Thornburg_2026}). Thankfully, development of frequency domain $m$-mode perturbation tools has accelerated over the past few years, with three different scalar perturbation projects being published recently~\cite{Vu_2026,Osburn_2022,Macedo_2024}. Naturally, advancing to the case of gravitational perturbations is a high priority. This work generalizes the approach of~\cite{Osburn_2022} to the case of Lorenz gauge metric perturbations. Initially we built a code designed to handle the full Kerr metric perturbation problem, but we were hampered by issues related to incorrect behavior near the horizon. To study and correct these near-horizon issues, in this work we have decided to focus on the simpler case of Schwarzschild metric perturbations calculated in terms of $m$-modes.

We calculate each $m$-mode of the first-order Lorenz gauge metric perturbation in the frequency domain by solving a system of ten coupled elliptic PDEs. Our discretized representation of the field equations follows from a second-order finite difference method, which produces a linear system describing the metric perturbation values at each $r$ and $\theta$ position. We developed code in \texttt{Mathematica} to build the associated sparse matrix and solve the linear system numerically.  We regularize the fields near the secondary via the effective source method, and for the first time we present the details of an associated Kerr puncture calculation that has supported certain gravitational perturbation results for years~\cite{Isoyama_2014}. A section is devoted to investigating obstacles we encountered near the horizon and then explaining our strategy to overcome them. We focus on circular orbital motion for simplicity and on radiative modes ($m\ne 0$) because they describe the type of outgoing flux that is the key ingredient for downstream second-order self-force models. We conclude by foreshadowing a more sophisticated and robust code for the full Kerr case based on \texttt{SpECTRE} that is in development.

\section{$m$-mode scheme in Lorenz gauge}

\subsection{Metric perturbation equations}

Consider a small compact object with mass $\mu$ orbiting a Schwarzschild black hole with mass $M$. To describe the gravitational dynamics, we introduce linear Schwarzschild metric perturbations
\begin{align}
\label{eq:gauge}
 \mathbf{g}_{\mu\nu} = g_{\mu\nu} + \epsilon \, h_{\mu\nu} + \mathcal{O}(\epsilon^2) \, ,
\qquad
 \epsilon \equiv \frac{\mu}{M} \ll 1 \, ,
\end{align}
where $g_{\mu\nu}$ is the Schwarzschild metric, $\epsilon$ is the small mass-ratio\footnote{It would be better to use the symmetric mass-ratio, but that is simple to achieve by re-interpreting these results.}, and $h_{\mu\nu}$ is the first-order metric perturbation. One way of calculating $h_{\mu\nu}$ that has certiain advantages for first- and second-order self-force applications is to adopt the Lorenz gauge condition
\begin{align}
\label{eq:gauge}
& g^{\alpha\mu} \nabla_\alpha\, \bar{h}_{\mu\nu} = 0 \, ,
\end{align}
where $\nabla_\alpha$ is the covariant derivative associated with $g_{\mu\nu}$ and $\bar{h}_{\mu\nu}$ is the trace-reversed metric perturbation
\begin{align}
&\bar{h}_{\mu\nu} = h_{\mu\nu}-\frac{1}{2} g_{\mu\nu} \, g^{\alpha\beta}\, h_{\alpha\beta} \, .
\end{align}
The Lorenz gauge has desirable features including a certain amount of regularity approaching point sources (which is especially useful for second-order calculations) and field equations with a relatively simple form
\begin{align}
\label{eq:fieldeqs}
& g^{\alpha\beta}\nabla_\alpha \nabla_\beta \,\bar{h}_{\mu\nu}+2R^{\alpha\;\beta}_{\;\,\mu\;\nu}\,\bar{h}_{\alpha\beta}= -16\pi \, T_{\mu\nu} \, ,
\end{align}
where $R^{\alpha\;\beta}_{\;\,\mu\;\nu}$ is the Riemann tensor associated with $g_{\mu\nu}$ and $T_{\mu\nu}$ is the stress energy tensor of the small body, which we represent as a point particle. Equation \eqref{eq:fieldeqs} follows from a linear perturbation of the Einstein tensor, and Eqs.~\eqref{eq:gauge} and \eqref{eq:fieldeqs} must both be satisfied for the Lorenz gauge metric perturbation to be a valid solution of the linearized Einstein field equations. Moving forward, we adopt Schwarzschild coordinates, $x^\alpha = (t, r, \theta, \phi)$:
\begin{align}
&g_{\mu\nu}\,dx^\mu dx^\nu = -\frac{\Delta}{r^2}dt^2+\frac{r^2}{\Delta}dr^2+r^2\left(d\theta^2+\sin^2\theta\,d\phi^2  \right) ,
\end{align}
where $\Delta \equiv r^2-2Mr$.
For simplicity, we consider the case where the small body follows a circular geodesic with position $x_p^\alpha$:
\begin{align}
& r_p = r_0\, , \qquad \theta_p = \frac{\pi}{2} \, , \qquad \phi_p = \Omega \, t \, , \qquad
\end{align}
where $r_0$ is the radius of the circular orbit and where
\begin{align}
\Omega = \sqrt{\frac{M}{r_0^3}}
\end{align}
is the angular speed that follows from standard timelike geodesic behavior.

Our calculation of the metric perturbation is based on separating the $\phi$ and $t$ variables by representing $\bar{h}_{\mu\nu}$ as a sum over $m$-modes
\begin{align}
\label{eq:modes}
\bar{h}_{\mu\nu}(t,r,\theta,\phi) = \sum_m \bar{h}^m_{\mu\nu}(r,\theta)\, e^{im(\phi-\Omega t)}\, ,
\end{align}
which shifts the focus to calculating how each $\bar{h}^m_{\mu\nu}$ depends on $r$ and $\theta$. This approach is designed to enable direct Kerr metric perturbation calculations without relying on metric reconstruction. For favorable numerical properties, it is convenient to introduce prefactors that depend on $r$ and $\theta$ (following Dolan and Barack~\cite{Dolan_2013})
\begin{align}
\label{eq:vec}
\vec{\psi}_m = \left(\begin{array}{c} 
\vphantom{\frac{r^2}{\Delta}}\psi_m^{0} \\ \vphantom{\frac{r^2}{\Delta}}\psi_m^{1} \\ \vphantom{\frac{r^2}{\Delta}}\psi_m^{2} \\ \vphantom{\frac{r^2}{\Delta}}\psi_m^{3} \\ \vphantom{\frac{r^2}{\Delta}}\psi_m^{4} \\ \vphantom{\frac{r^2}{\Delta}}\psi_m^{5} \\
\vphantom{\frac{r^2}{\Delta}}\psi_m^{6} \\ \vphantom{\frac{r^2}{\Delta}}\psi_m^{7} \\ \vphantom{\frac{r^2}{\Delta}}\psi_m^{8} \\ \vphantom{\frac{r^2}{\Delta}}\psi_m^{9}
\end{array}\right) = \left(\begin{array}{c}
r\,\bar{h}^m_{tt} \vphantom{\frac{r^2}{\Delta}}\\ \frac{\Delta}{r}\bar{h}^m_{tr} \vphantom{\frac{r^2}{\Delta}}\\ \bar{h}^m_{t\theta} \vphantom{\frac{r^2}{\Delta}}\\ 
\frac{1}{\sin{\theta}}\bar{h}^m_{t\phi} \vphantom{\frac{r^2}{\Delta}}\\ \frac{\Delta^2}{r^3}\bar{h}^m_{rr} \vphantom{\frac{r^2}{\Delta}}\\ \frac{\Delta}{r^2}\bar{h}^m_{r\theta} \vphantom{\frac{r^2}{\Delta}}\\ \frac{\Delta}{r^2\sin{\theta}}\bar{h}^m_{r\phi} \vphantom{\frac{r^2}{\Delta}}\\\;\; \frac{1}{r}\bar{h}^m_{\theta\theta} \vphantom{\frac{r^2}{\Delta}}\\ \frac{1}{r\sin{\theta}}\bar{h}^m_{\theta\phi} \vphantom{\frac{r^2}{\Delta}}\\ \frac{1}{r\sin^2\theta}\bar{h}^m_{\phi\phi} \vphantom{\frac{r^2}{\Delta}}
\end{array}\right) \, .
\end{align}
Among other things, these prefactors cause the numerical solutions to approach constant (or sometimes zero at the poles) amplitudes at all boundaries, see Sec.~\ref{sec:default} for a more detailed discussion. It can be shown that substitution of Eqs.~\eqref{eq:vec} and \eqref{eq:modes} into Eq.~\eqref{eq:fieldeqs} produces (after simplifications) the following elliptic PDEs governing each $m$-mode
\begin{align}
\label{eq:PDEs}
\left(\frac{\partial^2}{\partial r_*^2} + \frac{\Delta}{r^4} \frac{\partial^2}{\partial \theta^2}  + \mathbf{A}_m \frac{\partial}{\partial r_*}  + \mathbf{B}_m \frac{\partial}{\partial \theta}  + \mathbf{C}_m \right) \vec{\psi}_m = \vec{S}_m \, ,
\end{align}
where $\mathbf{A}_m$, $\mathbf{B}_m$, and $\mathbf{C}_m$ are $10\times 10$ matrices of known functions of $r$ and $\theta$ (these matrices are provided in Appendix~\ref{sec:pdes}), and $r_*$ is the tortoise coordinate defined by
\begin{align}
r_* = r + 2M \ln{\left(\frac{r}{2M}-1\right)} \, ,
\end{align}
which is a helpful radial coordinate to study gravitational wave propagation because $t\pm r_*$ is null.

Note that this work only considers $m \ne 0$ modes, which are the radiative modes (although the $m=\pm 1$ modes have a mixture of radiative and non-radiative content). This is partly because we are calculating fluxes as our primary numerical result (and $m=0$ does not contribute for circular motion), but also partly because of issues we encountered that are discussed in Section~\ref{sec:hor}.

\section{Effective source regularization}

We next consider the source for our equations, i.e. the right-hand-side of Eq.~\eqref{eq:PDEs}. Treating the small body as a point particle with Dirac-delta stress energy, the solution for the retarded field has a Coulomb-type divergence: $\bar{h}^{\rm ret}_{\mu\nu} \sim 1/\lambda$ where $\lambda$ is a formal order-counting parameter that counts powers of distance from the particle's worldline.
The corresponding $m$-modes of this field have a weaker logarithmic divergence: $\vec{\psi}_m \sim \log \lambda$.

As in Ref.~\cite{Vu_2026}\footnote{In this proof-of-principle calculation we did not find it necessary to include some of the sophisticated features (such as the use of Legendre-polynomial expressions for integrals or power series approximations near the worldline) added in Ref.~\cite{Vu_2026}. These will, however, be incorporated in an improved implementation in \texttt{SpECTRE}.}, we avoid having to handle this divergence directly by using the puncture/effective source method. We introduce a \textit{puncture} field $\bar{h}^{\cal P}_{\mu\nu}$ that locally captures the singular behavior. Subtracting the puncture field from the retarded field we get a \textit{residual} field:
\begin{align}
\label{eq:singular-regular}
\bar{h}_{\mu\nu}^{\cal{R}} = \bar{h}_{\mu\nu} - \bar{h}_{\mu\nu}^{\cal{P}}.
\end{align}
Substituting Eq.~\eqref{eq:singular-regular} into the linearised Einstein equation, Eq.~\eqref{eq:fieldeqs}, we obtain an equation for the residual field,
\begin{equation}
g^{\alpha\beta}\nabla_\alpha \nabla_\beta \,\bar{h}^{\cal R}_{\mu\nu}+2R^{\alpha\;\beta}_{\;\,\mu\;\nu}\,\bar{h}^{\cal R}_{\alpha\beta}
= S^{\rm eff}_{\mu\nu},
\end{equation}
where
\begin{equation}
   S^{\rm eff}_{\mu\nu} := -16\pi \, T_{\mu\nu} - g^{\alpha\beta}\nabla_\alpha \nabla_\beta \,\bar{h}^{\cal P}_{\mu\nu}+2R^{\alpha\;\beta}_{\;\,\mu\;\nu}\,\bar{h}^{\cal P}_{\alpha\beta}
\end{equation}
is the \textit{effective source}. At the level of $m$-modes we have the corresponding equation
\begin{align}
\label{eq:effectivePDEs}
\left(\frac{\partial^2}{\partial r_*^2} + \frac{\Delta}{r^4} \frac{\partial^2}{\partial \theta^2}  + \mathbf{A}_m \frac{\partial}{\partial r_*}  + \mathbf{B}_m \frac{\partial}{\partial \theta}  + \mathbf{C}_m \right) \vec{\psi}^{\cal R}_m = \vec{S}^{\rm eff}_m \, ,
\end{align}
with
\begin{align}
\label{eq:effectivePDEs}
\vec{S}^{\rm eff}_m &= \vec{S}_m \nonumber \\
& - \left(\frac{\partial^2}{\partial r_*^2} + \frac{\Delta}{r^4} \frac{\partial^2}{\partial \theta^2}  + \mathbf{A}_m \frac{\partial}{\partial r_*}  + \mathbf{B}_m \frac{\partial}{\partial \theta}  + \mathbf{C}_m \right) \vec{\psi}^{\cal P}_m.
\end{align}

We obtain an expression for the puncture field in the same way as described in Ref.~\cite{Thornburg_2017}, but adapted to the gravitational case. As in that reference, we adopt a fourth-order puncture (keeping four terms in the local series expansion, $\bar{h}_{\mu\nu}^{\cal{P}}\sim 1/\lambda + \lambda^0 + \lambda^1 +\lambda^2$), we use the ``Q-R'' scheme from Ref.~\cite{Wardell_2012}, and we apply the methods of Refs.~\cite{Heffernan_2012,Heffernan_2014} to obtain a four-dimensional puncture field given by
\begin{align}
\label{eq:puncture}
\bar{h}_{\mu\nu}^{\cal P}(x^\alpha; x^\alpha_p) = \frac{1}{\hat{\rho}^7} \sum_{i,j,k,l} A_{\mu\nu}^{ijkl}(x_p^\alpha, u^\alpha) \Delta r^i \Delta \theta^j Q^k R^l,
\end{align}
where $\Delta r := r-r_0$, $\Delta \theta := \theta - \pi/2$, $Q:=\sin (\Delta \varphi/2)$, $R:=\sin (\Delta \varphi)$, and
\begin{equation}
\label{eq:s2}
    \hat{\rho}^2 := \varrho^2 + z_c^2\, Q^2
\end{equation}
with
\begin{subequations}
\begin{align}
    \varrho^2 &:= \frac{r_p^2}{r_p^2-2M r_p+a^2} \Delta r^2 + r_p^2 \Delta \theta^2, \\
    z_c^2 &:= 4 \left({\cal L}^2 + r_p^2 + a^2 + \frac{2M a^2}{r_p}\right).
\end{align}
\end{subequations}
The sum here is over all $i$, $j$, $k$ and $l$ such that $6\le i+j+k+l \le 9$ and $l \in \{0,1\}$.
Note that the coefficients $A_{\mu\nu}^{ijkl}$ are functions of the particle position and four-velocity only, and that for our case of circular equatorial orbits we have that $A_{\mu\nu}^{ijkl}=0$ in several cases:
\begin{itemize}
    \item If neither or both of $\mu$ and $\nu$ are $\theta$ components and either $j$ or $k$ is odd;
    \item If only one of $\mu$ or $\nu$ is a $\theta$ component and either $j$ is even or $k$ is odd;
    \item If neither or both of $\mu$ and $\nu$ are $t$ or $\phi$ components and $l=1$;
    \item If only one of $\mu$ or $\nu$ is a $t$ or $\phi$ component and $l=0$.
\end{itemize}
These are a consequence of the equatorial-plane and time-reversal symmetries of the circular, equatorial orbit problem.

Our puncture field comprises 10 Boyer-Lindquist tensor components of the metric perturbation. Each component is in exactly the same form as for the scalar field in Ref.~\cite{Thornburg_2017}, so we obtain a decomposition into $m$-modes in the same way. In particular, we use the expressions in Sec. II G of Ref.~\cite{Thornburg_2017} to write the $m$-modes of the puncture in terms of complete elliptic integrals whose arguments are functions of $\Delta \theta$, $\Delta r$ and the worldline position and four-velocity. We implemented these final expressions, along with expression for the effective source (which can similarly be written in terms of complete elliptic integrals and the coefficients $A_{\mu\nu}^{ijkl}$) in the \texttt{GravitationalEffectiveSource} C code \cite{Wardell_github}. This code was used to generate data on the same grid as required by the \texttt{Mathematica} field equation numerical solver. As described in the next section, we require this data only within a worldtube around the particle, along with boundary data consisting of values of the puncture field in the vicinity of the worldtube.

\section{Numerical methods}

\label{sec:num}

\subsection{Finite difference scheme}

\begin{figure}
\includegraphics[width=3.4in]{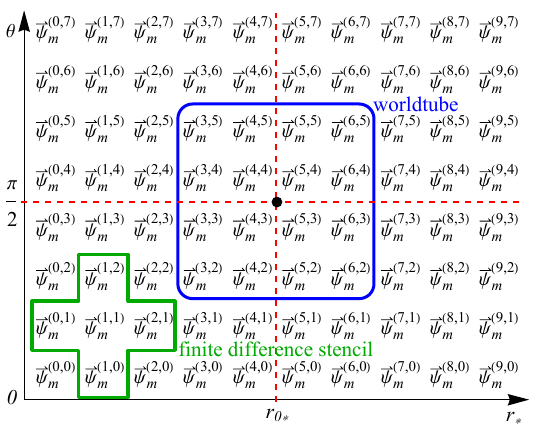}
\caption{\label{fig:discretize} Our rectangular discretization of the $r_*$-$\theta$ domain is shown. The black dot represents the position of the smaller binary component. We adopt a uniform grid spacing in both $\Delta r_*$ and $\Delta\theta$. Outside the worldtube we calculate the retarded field at each grid point, while inside the worldtube we calculate the residual field.}
\end{figure}

To numerically solve the system of 10 coupled elliptic PDEs of Eq.~\eqref{eq:effectivePDEs} we apply the same finite-difference strategy as Osburn and Nishimura~\cite{Osburn_2022}, which we summarize here. Our discretization imposes uniform grid spacings $\Delta r_*$ and $\Delta\theta$, and we assign $(i,j)$ indices to represent $r_*$ and $\theta$ positions respectively: $\vec{\psi}_m^{(i,j)}\equiv\vec{\psi}_m(r_{*i},\theta_j)$, see Fig.~\ref{fig:discretize}. Applying this discretization to the PDEs involves representing the derivatives with finite difference approximations
\begin{align}
\frac{\partial^2\vec{\psi}_m}{\partial r_*^2} &\simeq \frac{\vec{\psi}_m^{(i+1,j)}-2\vec{\psi}_m^{(i,j)}+\vec{\psi}_m^{(i-1,j)}}{\Delta r_*^2} \, ,
\\
\frac{\partial^2\vec{\psi}_m}{\partial \theta^2} &\simeq \frac{\vec{\psi}_m^{(i,j+1)}-2\vec{\psi}_m^{(i,j)}+\vec{\psi}_m^{(i,j-1)}}{\Delta\theta^2} \, ,
\\
\frac{\partial\vec{\psi}_m}{\partial r_*} &\simeq \frac{\vec{\psi}_m^{(i+1,j)}-\vec{\psi}_m^{(i-1,j)}}{2\Delta r_*} \, ,
\\
\frac{\partial\vec{\psi}_m}{\partial\theta} &\simeq \frac{\vec{\psi}_m^{(i,j+1)}-\vec{\psi}_m^{(i,j-1)}}{2\Delta\theta} \, ,
\end{align}
where we use the standard five-point second-order finite difference stencil. Inserting this logic into the differential equations throughout the grid produces a system of coupled linear algebra equations. It is convenient to represent this linear system in matrix form, and we built a \texttt{Mathematica} code to construct this sparse matrix.

As is standard in the effective source approach \cite{Wardell_2012}, we restrict the support of our puncture field to a worldtube around the particle. The source $\vec{S}^{\rm eff}_m$ is then only non-zero within the worldtube and there is a step-function jump in the numerical solution at the edges of the worldtube. At the level of the linear system, this translates to a right-hand-side source vector that is only non-zero for components corresponding to points inside the worldtube. The jump discontinuities are imposed by transforming between $\vec{\psi}_m$ and $\vec{\psi}_m^\mathcal{R}$ according to Eq.~\eqref{eq:singular-regular} when the finite difference stencil intersects the worldtube (this is the same strategy introduced by Osburn and Nishimura~\cite{Osburn_2022}).

For each $m$-mode, the associated linear system is solved with the built-in \texttt{Mathematica} function \texttt{LinearSolve} with the ``Pardiso" option. Despite efforts to optimize the code, our \texttt{Mathematica} implementation is rather slow to construct these matrices ($\sim 3$ hours per $m$-mode when the resolution is high), and constructing the matrix is significantly slower than solving the linear system afterwards. 
Inefficiencies such as these have caused accurate solutions to be rather costly (see Sec.~\ref{sec:flux} for resolution convergence results) and have motivated a more sophisticated implementation based on the \texttt{SpECTRE} numerical relativity code (see \cite{Vu_2026} for initial progress with a scalar-field implementation).

\subsection{Boundary conditions}

We follow Dolan and Barack~\cite{Dolan_2013} by enforcing regularity of the metric perturbation at the poles. This is achieved by requiring continuity and differentiability in a locally Cartesian coordinate system, which implies an associated set of regularity conditions to impose on the Schwarzschild coordinate components of $\bar{h}_{\mu\nu}$, which finally leads to $\theta=0$ and $\theta=\pi$ boundary conditions for $\vec{\psi}_m$. Care must be taken when approaching the poles because the behavior depends on the $m$-mode eigenfunctions, $e^{im\phi}$, and the prefactors relating $\bar{h}^m_{\mu\nu}$ to $\vec{\psi}_m$ (which involve functions of $\theta$, see Eq.~\eqref{eq:vec}). These considerations cause the type of boundary condition to depend on $m$:
\begin{align}
\label{eq:thA}
\psi^{0,1,4,7,8,9}_{m=1}\Big|_{\theta=0,\pi} &= 0 \, ,
\\
\frac{\partial \psi^{2,3,5,6}_{m=1}}{\partial \theta}\bigg|_{\theta=0,\pi} &= 0 \, ,
\\
\psi^{0,1,2,3,4,5,6}_{m=2}\Big|_{\theta=0,\pi} &= 0 \, ,
\\
\frac{\partial \psi^{7,8,9}_{m=2}}{\partial \theta}\bigg|_{\theta=0,\pi} &= 0 \, ,
\\
\vec{\psi}_{m>2}\Big|_{\theta=0,\pi} &= 0 \, .
\label{eq:thB}
\end{align}
Therefore, some components satisfy a Dirichlet condition at the poles, and other components satisfy a Neumann condition at the poles. For cases involving a Neumann condition, we use a second-order one-sided finite difference representation of the $\theta$ derivatives
\begin{align}
\label{eq:thetaBC}
\frac{\partial \psi}{\partial \theta}\bigg|_{\theta=0} &\simeq \frac{-3 \psi\big|_{\theta=0} + 4 \psi\big|_{\theta = \Delta \theta} -\psi\big|_{\theta=2\Delta \theta}}{2\Delta\theta} ,
\\
\frac{\partial \psi}{\partial \theta}\bigg|_{\theta=\pi}  &\simeq \frac{3 \psi\big|_{\theta=\pi} - 4 \psi\big|_{\theta = \pi - \Delta \theta} + \psi\big|_{\theta=\pi - 2\Delta \theta}}{2\Delta\theta} \, . \notag
\end{align}

The radial domain spans from the near-horizon region at a sufficiently negative $r_*^\text{min}$ out towards distant observers at a sufficiently positive $r_*^\text{max}$ (we explore optimal values for both parameters). The boundary conditions imposed at $r_*^\text{min}$ and $r_*^\text{max}$ influence the propagation direction of asymptotic radiation, which is essential for producing the retarded solution. 

At $r_*^\text{max}$, we follow Osburn and Nishimura~\cite{Osburn_2022} by identifying the following boundary conditions that enforce an outgoing direction of radiation propagation
\begin{subequations}
\label{eq:BCs}
\begin{align}
\label{eq:BC1}
& \frac{\partial \vec{\psi}_{m}}{\partial r_*} -im\Omega \, \vec{\psi}_{m} = \mathcal{O}\Big(\frac{1}{r_*^{\text{max}^{\scriptstyle 2}}}\Big) \, ,
\\
\label{eq:BC2}
& \frac{\partial^2 \vec{\psi}_{m}}{\partial r_*^2} -2 im\Omega \frac{\partial \vec{\psi}_{m}}{\partial r_*} - m^2\Omega^2 \vec{\psi}_{m} = \mathcal{O}\Big(\frac{1}{r_*^{\text{max}^{\scriptstyle 3}}}\Big) \, ,
\\
& \frac{\partial^3 \vec{\psi}_{m}}{\partial r_*^3} -3 im\Omega \frac{\partial^2 \vec{\psi}_{m}}{\partial r_*^2} -3m^2\Omega^2\frac{\partial \vec{\psi}_{m}}{\partial r_*} +i m^3\Omega^3 \vec{\psi}_{m} \notag
\\&\qquad\qquad\qquad\qquad\qquad\qquad\qquad =\mathcal{O}\Big(\frac{1}{r_*^{\text{max}^{\scriptstyle 4}}}\Big) \, .
\label{eq:BC3}
\end{align}
\end{subequations}
All three above conditions are only approximate because the solution is not exactly a sinusoidal wave until $r_* \simeq \infty$. Equation~\eqref{eq:BC1} is the outgoing Sommerfeld condition and Eqs.~\eqref{eq:BC2} and~\eqref{eq:BC3} are outgoing frequency domain implementations of second- and third-order Bayliss-Turkel radiation conditions \cite{Bayliss_1980}. One-sided finite differences similar to Eq.~\eqref{eq:thetaBC} are used to represent these radial derivatives at boundaries. Our code adopts Eq.~\eqref{eq:BC3} at $r_*^\text{max}\simeq 500 M$, see Osburn and Nishimura~\cite{Osburn_2022} for full details.

Usage of the $r_*$ coordinate should enable logic similar to Eq.~\eqref{eq:BC1} for boundary conditions near the black-hole horizon
by implementing Eq.~\eqref{eq:BC1} at $r_*^\text{min}$, but with a sign change to select propagation towards the horizon. Prior work in the scalar-field case successfully applied this approach and obtained the correct retarded solution, but early attempts at using the same logic (a downgoing Sommerfeld condition with $r_*^\text{min}\simeq -50M$) failed in the gravitational case. Section~\ref{sec:hor} demonstrates and investigates this near-horizon boundary condition issue in detail, and explains how we mitigated those issues to produce our results.

\section{Issues near the horizon}

\label{sec:hor}

Up to this point, everything we have described has been a straightforward extension of the scalar-field calculation detailed in Ref.~\cite{Osburn_2022}. Other than the increased complexity of the equations in the gravitational case, the fundamental method has been identical.

When solving the gravitational problem, we did, however, encounter one problem that did not appear in the scalar-field case: we found that our numerical solutions did not exhibit the expected behavior towards the black-hole horizon. Comparison against the time-domain results of Dolan and Barack \cite{Dolan_2013} further revealed that our numerical solution was close to, but differed from the correct retarded solution, with the difference most pronounced towards the horizon. After a detailed analysis, we identified the problem as arising from nonphysical incoming-wave solutions near the horizon. These are not fully excluded by our use of (a near-horizon version of) Eq.~\eqref{eq:BC1}, which is only an approximation to the true behavior (both because it is evaluated at finite $r_*^\text{min}$ and because it involves finite-difference approximations to derivatives). In the rest of this section we analyze this problem in full detail and provide work-around solutions.

\subsection{Nonphysical numerical solution}
\label{sec:default}

The problem of Lorenz gauge Schwarzschild metric perturbations has been solved with various methods, and the behavior of the solution near the horizon is well understood. The correct retarded solution will have the ingoing Eddington-Finkelstein components of $\bar{h}_{\mu\nu}$ approach constant amplitudes near the horizon; this fact motivated the prefactors in Eq.~\eqref{eq:vec}, which enable consistency between $\vec{\psi}_m$ approaching constant amplitudes near the horizon and the Eddington-Finkelstein components of the metric perturbation approaching constant amplitudes near the horizon. In the early stages of this project, we hypothesized that imposing approximate downgoing (at $r_*=r_*^\text{min}$) and outgoing (at $r=r_*^\text{max}$) boundary conditions would be sufficient to identify the unique retarded solution of Eq.~\eqref{eq:fieldeqs}; however, we observed that the numerical solutions (following the methods of Sec.~\ref{sec:num}) tend to decay exponentially to zero approaching the horizon, which is inconsistent with the correct behavior of the retarded solution. 

\begin{figure}
\includegraphics[width=3.4in]{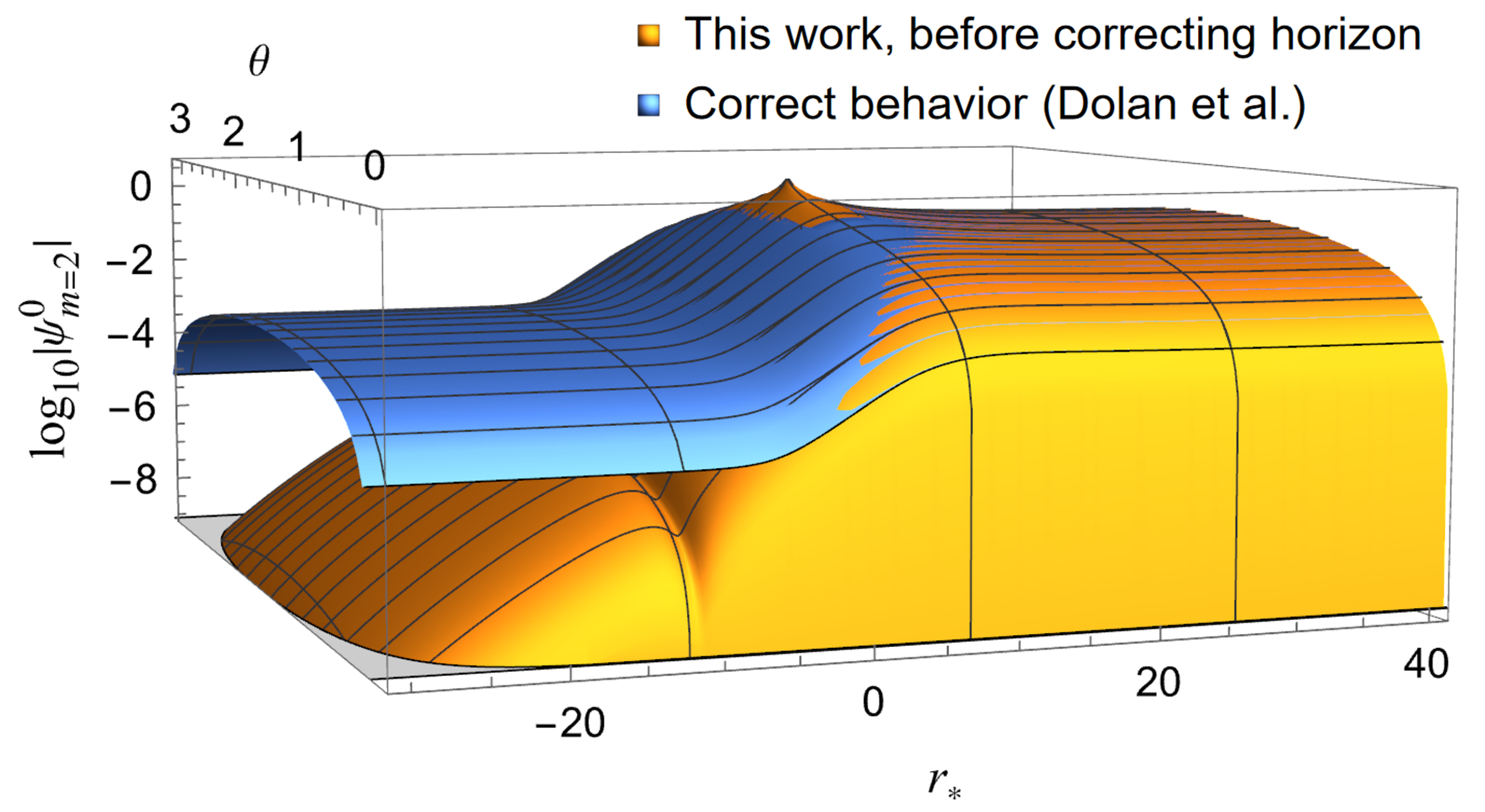}
\caption{\label{fig:m2issue} Near horizon exponential decay of our early numerical results are shown for the case with $m=2$ and $r_0 = 6M$. The log of the absolute value of $\psi^0_m$ is shown to illustrate this incorrect behavior. We are using the correct time-domain results from Dolan and Barack~\cite{Dolan_2013} for comparison. Notice that the correct $m=2$ behavior involves near-horizon waves that approach a constant amplitude approximately two orders of magnitude smaller than the large radius amplitude. We speculate that having a sufficiently small horizon amplitude for the correct $m=2$ data explains why our early data that decays near the horizon seems to agree at large radii despite the fact that incorrect horizon behavior should hurt the global solution.}
\end{figure}

Prior to displaying corrected final results (which are found in Sec.~\ref{sec:res}), it is useful here to illustrate these near-horizon issues by showing some earlier numerical results with the incorrect horizon behavior that follows from a naive approach to the near-horizon boundary condition. Figure~\ref{fig:m2issue} shows earlier numerical results that decay exponentially approaching the horizon for the case with $m=2$ and $r_0=6M$. Notice that the correct $m=2$ behavior involves near-horizon waves that approach a constant amplitude approximately two orders of magnitude smaller than the large radius amplitude. We speculate that having a sufficiently small horizon amplitude for the correct $m=2$ behavior explains why our early data that decays near the horizon seems to agree at large radii despite the fact that incorrect horizon behavior should hurt the global solution. Although we are illustrating the near-horizon issues by showing exponential decay of $\psi^0_m$, overall, we observe that naive horizon boundary conditions lead to incorrect exponential decay of $\psi^{0,1,2,3,4,5,6}_m$ while $\psi^{7,8,9}_m$ approach constant amplitudes at the horizon with roughly correct behavior.

\begin{figure}
\includegraphics[width=3.4in]{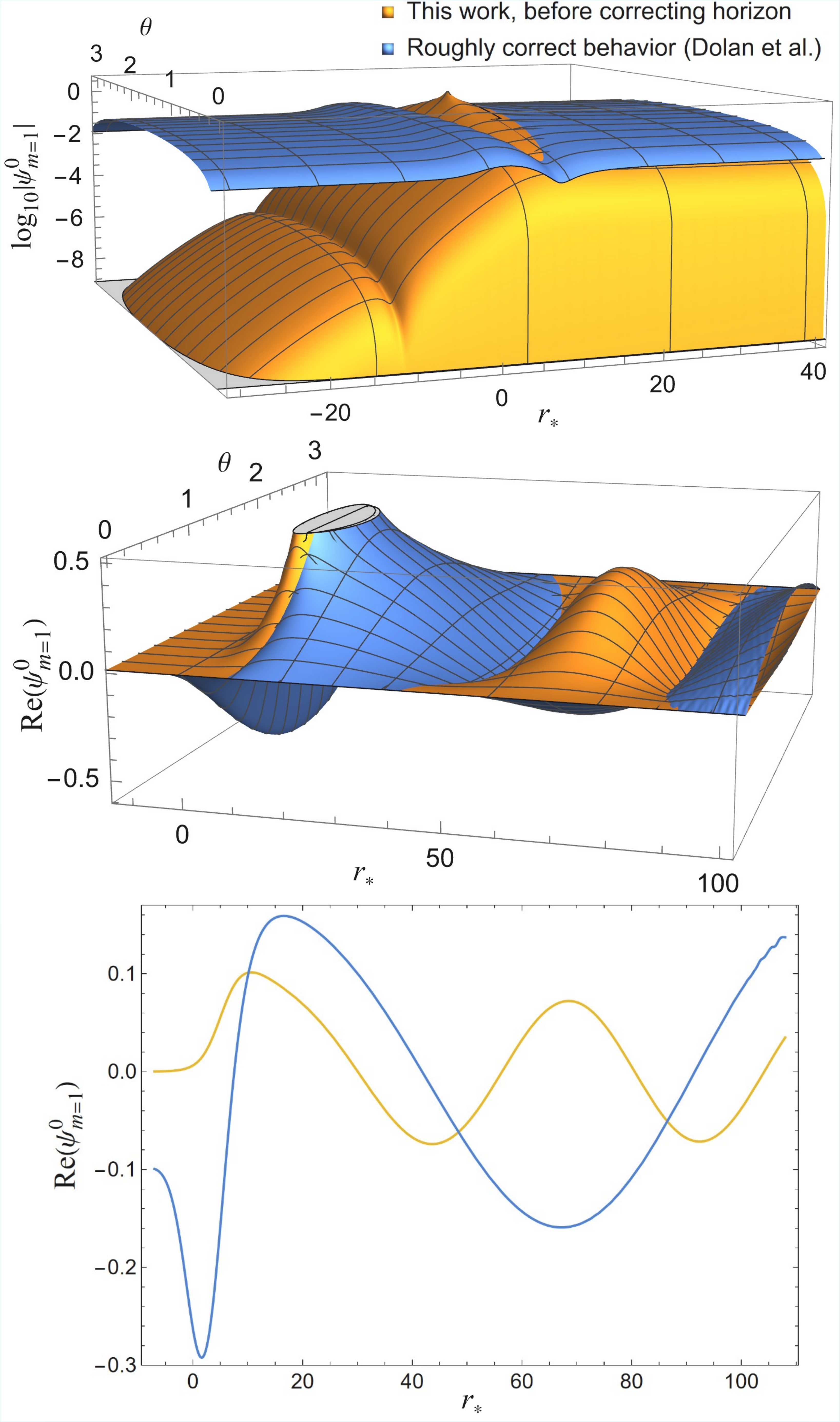}
\caption{\label{fig:m1issue} Behavior of our early numerical results are shown for the case with $m=1$ and $r_0 = 6M$. The top plot illustrates the near horizon exponential decay issue by showing the log of the absolute value of $\psi^0_m$ (like Fig.~\ref{fig:m2issue}), the middle plot shows the real part of $\psi^0_m$, and the bottom plot is a $\theta=0.5 \; \text{radian}$ cross-section of the real part of $\psi^0_m$. Notice that, unlike the $m\ge 2$ cases, our early $m=1$ results disagreed even away from the horizon; we believe this is because the correct $m=1$ horizon amplitude is not small compared to the large radius amplitude, which means that an amount of relative error at the horizon will lead to approximately the same amount of relative error everywhere. For this reason, our efforts to correct the near-horizon behavior will use the more sensitive $m=1$ case to validate the success of potential near-horizon boundary strategies. Note that, unlike the $m\ge 2$ cases, the time-domain results that have been shared~\cite{Dolan_2013} for comparison are noisier for the $m=1$ case (notice the large radius noise in the middle and bottom plots); this is due to a separate issue that affects the $m=1$ mode in the time-domain approach, but despite the noise the time-domain comparison results are roughly correct (the noise range overlaps the correct behavior). Also note that work is underway to overcome Lorenz gauge time-domain issues~\cite{Thornburg_2026}.}
\end{figure}

The $m=1$ case even more clearly illustrates these near-horizon issues. Figure~\ref{fig:m1issue} shows earlier numerical results that decay exponentially approaching the horizon for the case with $m=1$ and $r_0=6M$. Notice that, unlike the $m\ge 2$ cases, our early $m=1$ results disagreed even away from the horizon; we believe this is because the correct $m=1$ horizon amplitude is not small compared to the large radius amplitude, which means that an amount of relative error at the horizon will lead to approximately the same amount of relative error everywhere. For this reason, our efforts to correct the near-horizon behavior will focus on the sensitive $m=1$ case to validate the success of potential near-horizon boundary strategies.

\subsection{Near-horizon analysis}
\label{sec:decay}

Although analytical predictions of the near-horizon behavior of the global inhomogeneous solution are not readily available (this is why numerical methods are needed in the first place), we are able to analytically predict the near-horizon behavior of individual homogeneous solutions to Eq.~\eqref{eq:PDEs} in a way that generalizes to Kerr spacetime (which is the eventual goal). Away from the source, the inhomogeneous solution is a linear combination of homogeneous solutions, so identifying the behavior of a complete set of homogeneous solutions near the horizon is the key to understanding issues there.

With the field equations represented in terms of $r_*$ derivatives, the leading-order near-horizon behavior is obtained by substituting $r=2M$ into Eq.~\eqref{eq:PDEs}. Note that the coefficients of the first and second $\theta$ derivatives vanish at the horizon
\begin{align}
& \lim_{r\rightarrow 2M} \frac{\Delta}{r^4} = 0 \, ,
\\
& \lim_{r\rightarrow 2M} \mathbf{B}_m = \mathbf{0} \, ,
\end{align}
where $\mathbf{0}$ is a $10 \times 10$ matrix of zeros. Therefore, a system of ODEs with $r_*$ derivatives governs the behavior there
\begin{align}
\label{eq:horeqs}
\left(\frac{\partial^2}{\partial r_*^2}   + \mathbf{A}_m \big|_{r=2M} \frac{\partial}{\partial r_*}  + \mathbf{C}_m \big|_{r=2M} \right) \vec{\psi}_m = 0 \, ,
\end{align}
where $\mathbf{A}_m$ and $\mathbf{C}_m$ become matrices that are approximately independent of $r$ when $r_*$ becomes very negative. These ODEs can be written in first-order form
\begin{align}
\label{eq:ODEs}
\frac{\partial}{\partial r_*} \left(\begin{array}{c} 
\vec{\psi}_m \\\,\\ \frac{\partial}{\partial r_*} \vec{\psi}_m
\end{array}\right) = \left(\begin{array}{cc}
\mathbf{0} & \mathbf{I} \\\,&\,\\ -\mathbf{C}_m & -\mathbf{A}_m
\end{array}\right)\left(\begin{array}{c} 
\vec{\psi}_m \\\,\\ \frac{\partial}{\partial r_*} \vec{\psi}_m
\end{array}\right) \, ,
\end{align}
where $\mathbf{I}$ is the $10\times 10$ identity matrix. Equation~\eqref{eq:ODEs} is a linear system of 20 first-order coupled ODEs whose 20 independent homogeneous solutions are accessible via the eigenvalues and eigenvectors of the $20\times 20$ matrix. The form of the homogeneous solutions is:
\begin{align}
\label{eq:homo}
\left(\begin{array}{c} 
\vec{\psi}_m \\\,\\ \frac{\partial}{\partial r_*} \vec{\psi}_m
\end{array}\right) = \left(\begin{array}{c} 
\vec{v} \\\,\\ \lambda \, \vec{v}
\end{array}\right) \, e^{\lambda r_*}\, ,
\end{align}
where $\lambda$ represents one of the 20 eigenvalues that describe the $r_*$ exponent (which encodes radial dependence near the horizon) and $\vec{v}$ represents an associated eigenvector. General solutions near the horizon could be constructed by forming linear combinations of the independent homogeneous solutions with coefficients that can be arbitrary functions of $\theta$.

A full eigenvector analysis is presented in Table~\ref{tab:eigenvectors} of Appendix~\ref{sec:eigenvecs}, but here it is sufficient for our discussion to focus solely on the eigenvalues, which are shown in Table~\ref{tab:eigenvals}. Correct near-horizon behavior requires the following two features: first, $\vec{\psi}_m$ must satisfy the four Lorenz gauge conditions of Eq.~\eqref{eq:gauge}; second, $\vec{\psi}_m$ must propagate in the downgoing direction by having $Im(\lambda)<0$. According to Table~\ref{tab:eigenvals}, the first feature invalidates 8 of the near-horizon solutions and the second feature further invalidates 6 solutions. Therefore, the correct near-horizon behavior involves a linear combination of the remaining 6 solutions who all share the eigenvalue $\lambda=-i m \Omega$. Note that $Re(\lambda) \ge 0$ for all 20 solutions; this implies that, if a solution does not approach a constant amplitude at the horizon, it can be thought of as either growing exponentially as $r_*$ increases or, equivalently, decaying exponentially as $r_*$ decreases. Because the early numerical results trend towards zero, we have largely adopted the ``decays exponentially as $r_*$ decreases" viewpoint. However, the ``grows exponentially as $r_*$ increases" viewpoint may provide some insight into why our early numerical results consistently favor zero at the horizon: if small artifacts such as finite difference errors were to uniformly excite all 20 solutions numerically, exponential growth in the increasing $r_*$ direction would cause the $Re(\lambda) > 0$ solutions to dominate while also having their accumulated global maximum be constrained by the size of the source. In other words, growing exponentially with a ceiling set by inhomogeneous features is only possible if the starting value near the horizon is very close to zero (this type of behavior has been observed in similar contexts~\cite{Osburn_2014}).

\begin{table}
\caption{\label{tab:eigenvals}
Eigenvalues describing the 20 independent near-horizon homogeneous solutions according to Eq.~\eqref{eq:homo} are shown. The ``degeneracy" tells how many different eigenvectors share that eigenvalue. The first row is consistent with the known correct behavior of the inhomogeneous solution. The explanation for whether or not solutions satisfy the Lorenz gauge conditions requires further information involving the associated eigenvectors, see Appendix~\ref{sec:eigenvecs} for full details. 
}
\begin{ruledtabular}
\begin{tabular}{c|c|c}
 Eigenvalue: $\lambda$ $\;$ & Degeneracy $\;\;$ & Satisfies Lorenz gauge? $\;$ $\displaystyle \vphantom{\frac{a}{b}}$ \\ 
 \hline
 $-i m \Omega \;$ (correct) & $\times 6 \;\;$ & yes $\;$ \\ 
 $\frac{1}{2M}-i m \Omega \;$ & $\times 3 \;\;$ & no $\;$ \\  
 $\frac{1}{M}-i m \Omega \;$ & $\times 1 \;\;$ & no $\;$ \\  
 $+i m \Omega \;$ & $\times 6 \;\;$ & yes $\;$ \\  
 $\frac{1}{2M}+i m \Omega \;$ & $\times 3 \;\;$ & no $\;$ \\  
 $\frac{1}{M}+i m \Omega \;$ & $\times 1 \;\;$ & no $\;$ \\ 
\end{tabular}
\end{ruledtabular}
\end{table}

\subsection{Improved near-horizon boundary conditions}

The correct near-horizon behavior of the retarded solution and the possible behavior of the 20 homogeneous solutions were not unexpected; however, the failure of our early results to follow the correct behavior was unexpected (by us) because our early naive boundary condition was motivated by the same eigenvalue analysis we just used to illustrate the issue. Our naive near-horizon boundary condition,
\begin{align}
\label{eq:naive}
& \frac{\partial \vec{\psi}^\text{naive}_{m}}{\partial r_*} + im\Omega \, \vec{\psi}^\text{naive}_{m} = 0 \, ,
\end{align}
was designed specifically to select the correct eigenvalue $\lambda = -im\Omega$; indeed, the correct retarded solution satisfies Eq.~\eqref{eq:naive}. Unfortunately, this condition was not sufficient to exclude contributions from incorrect homogeneous solutions because $\vec{\psi}_m=0$ satisfies Eq.~\eqref{eq:naive} equally well, and if zero satisfies the boundary condition then the problematic solutions that effectively decay to zero may be permitted numerically (note that $\vec{\psi}_m=0$ also satisfies the the large $r$ boundary condition, but that is not problematic because no unwanted solutions approach zero there). To ameliorate this problem, we have investigated two separate strategies to improve the near-horizon boundary conditions.

Our first strategy to improve the near-horizon boundary behavior is to identify conditions that can be applied further from the horizon without losing accuracy. We hypothesized that excluding as much of the near-horizon regime from the numerical domain as possible would help by avoiding the region where the issue is present. However, Eq.~\eqref{eq:naive} is only valid when $r_*$ is very negative, so we need to derive new boundary conditions with an increased range of validity. Consider a generic near-horizon expansion of $\vec{\psi}_m$
\begin{align}
\label{eq:horex}
\vec{\psi}_m = e^{-im\Omega r_*}\Big( \vec{f}(\theta) + \vec{g}(\theta)\,\Delta + \vec{h}(\theta)\,\Delta^2 + \ldots \Big) ,
\end{align}
where $\vec{f}$, $\vec{g}$, and $\vec{h}$ (and higher order coefficients) are functions of $\theta$ that in practice would be determined only after numerically finding the inhomogeneous solution. Because derivatives of $\vec{\psi}_m$ follow a similar pattern
\begin{align}
&\frac{\partial \vec{\psi}_m}{\partial r_*} = e^{-im\Omega r_*}\bigg( -im\Omega \vec{f} +\Big(\frac{1}{2M}-im\Omega \Big)\vec{g}\, \Delta + \ldots \bigg) ,
\end{align}
we can form linear combinations of $\vec{\psi}_m$ and its $r_*$ derivatives to derive boundary conditions that cancel errors that are present farther from the horizon. Each higher derivative of Eq.~\eqref{eq:horex} included in the linear combination can achieve cancellation of the next nonvanishing lowest-order residual term. The following three linear combinations achieve successively higher order error cancellations
\begin{subequations}
\begin{align}
& \frac{\partial \vec{\psi}_{m}}{\partial r_*} + im\Omega \, \vec{\psi}_{m} = \mathcal{O}(\Delta) \, ,
\\
& \frac{\partial^2 \vec{\psi}_{m}}{\partial r_*^2} + \left(2im\Omega-\frac{1}{2M} \right)\frac{\partial \vec{\psi}_{m}}{\partial r_*}  - \left(m^2\Omega^2+\frac{im\Omega}{2M} \right)\vec{\psi}_{m}  \notag
\\&\qquad\qquad\qquad\qquad\qquad\qquad\qquad\;\;\;= \mathcal{O}(\Delta^2) \, , \label{eq:horbc0}
\\
& \frac{\partial^3 \vec{\psi}_{m}}{\partial r_*^3} + \left( 3im\Omega-\frac{3}{2M} \right) \frac{\partial^2 \vec{\psi}_{m}}{\partial r_*^2} \notag
\\&\qquad - \left( 3m^2\Omega^2 +\frac{3im\Omega}{M} -\frac{1}{2M^2} \right)\frac{\partial \vec{\psi}_{m}}{\partial r_*} \label{eq:horbc}
\\&\qquad\qquad - \left( i m^3\Omega^3 - \frac{3m^2\Omega^2}{2M} - \frac{im\Omega}{2M^2} \right) \vec{\psi}_{m}  = \mathcal{O}(\Delta^3) \, .  \notag
\end{align}
\end{subequations}
Because associated numerical implementations represent each residual as zero, it is apparent that minimizing the size of the residual improves the accuracy of the solution associated with a certain boundary condition. Note the similarity between Eq.~\eqref{eq:horbc} and Eq.~\eqref{eq:BC3}; this is because Eqs.~\eqref{eq:horbc0}-\eqref{eq:horbc} are generalizations of the frequency-domain Bayliss-Turkel~\cite{Bayliss_1980} approach to the near-horizon regime (which, as far as we know, is a novel result). We experimented with versions of Eq.~\eqref{eq:horbc} that include as many as eight derivatives, which could (depending on error coefficients) produce a residuals of size $\simeq 10^{-14}$ at boundary positions as far from the horizon as $r_*^\text{min}\simeq-6M$. This strategy alone was not sufficient to correct the near-horizon issues, but it is a valuable ingredient when combined with the next strategy.

\begin{figure}
\includegraphics[width=3.4in]{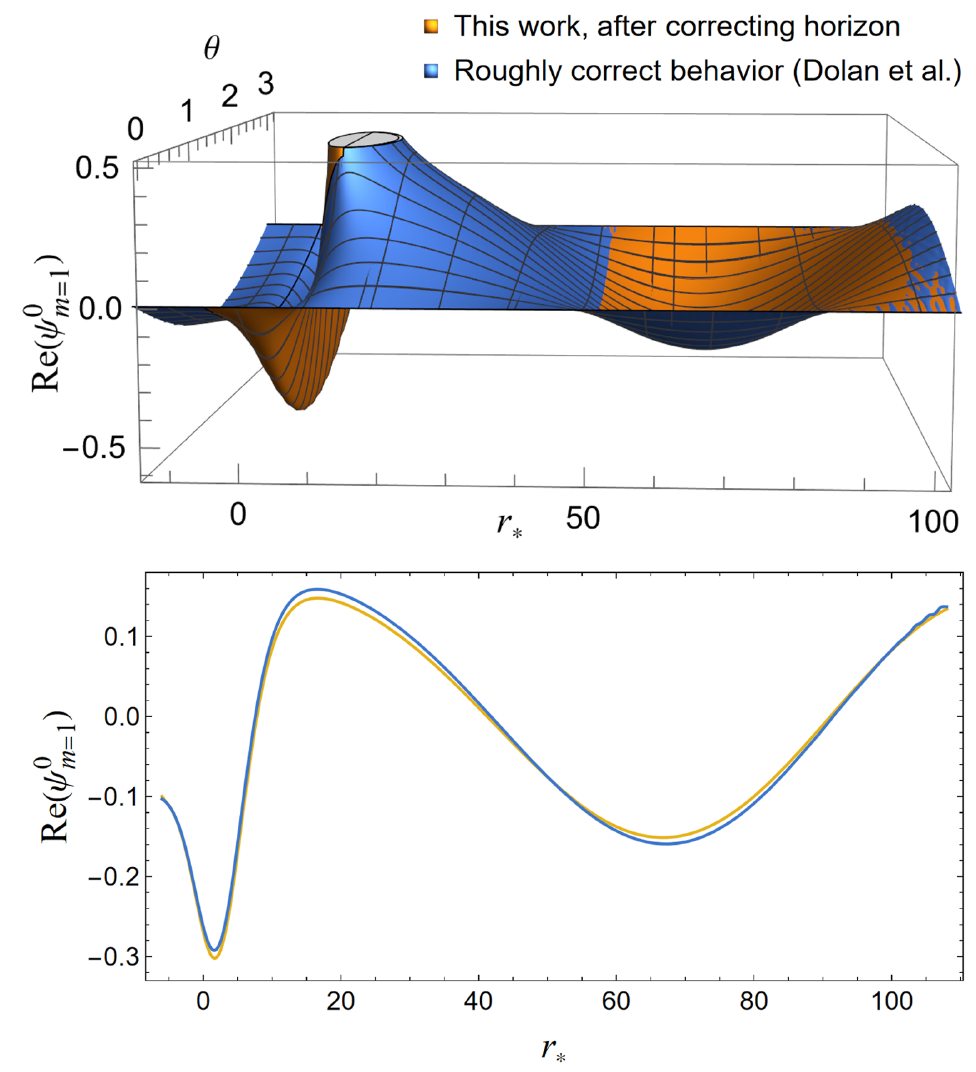}
\caption{\label{fig:m1correct} Behavior of our corrected numerical results are shown for the case with $m=1$ and $r_0 = 6M$. The top plot shows the real part of $\psi^0_m$ and the bottom plot is a $\theta=0.5 \; \text{radian}$ cross-section of the real part of $\psi^0_m$. Notice that, compared to Fig.~\ref{fig:m1issue}, our strategies to correct the near-horizon behavior have produced large improvements for the most sensitive $m=1$ mode (and our $m\ge 2$ results have much smaller overall errors than that).}
\end{figure}

Our second strategy to improve the near-horizon boundary behavior is to impose the Lorenz gauge conditions as a boundary condition at $r_*^\text{min}$. Initially, we tried using the near-horizon limit of the Lorenz gauge conditions imposed closer to the horizon ($r_*^\text{min}\simeq-50M$). Observing that the limiting behavior was insufficient, we adopted a strategy where Eq.~\eqref{eq:gauge} is imposed without analytic approximations; the implementation was interesting because now the $r_*^\text{min}$ boundary condition involves the $\theta$ derivatives present in the full gauge conditions, but it does not seem to cause any issues.

This clarifies our combined strategy for the near-horizon boundary condition: we choose a boundary position relatively far from the horizon ($r_*^\text{min}\simeq-6M$) to avoid the problematic region, introduce four boundary conditions that impose the four Lorenz gauge conditions, and then use a higher-order version of Eq.~\eqref{eq:horbc} applied to $\psi_m^{4,5,6,7,8,9}$ (six of the metric perturbation components) to assemble a complete set of ten near-horizon boundary conditions. Figure~\ref{fig:m1correct} demonstrates that, even for the $m=1$ case that is most sensitive to the near-horizon boundary condition, our new strategies have achieved results that are roughly within the noise of the time-domain comparison code (and our $m\ge 2$ results are much more accurate, see Sec.~\ref{sec:flux} for numbers).

\section{Results}
\label{sec:res}

\subsection{Metric perturbation}

\begin{figure*}
\includegraphics[width=\textwidth]{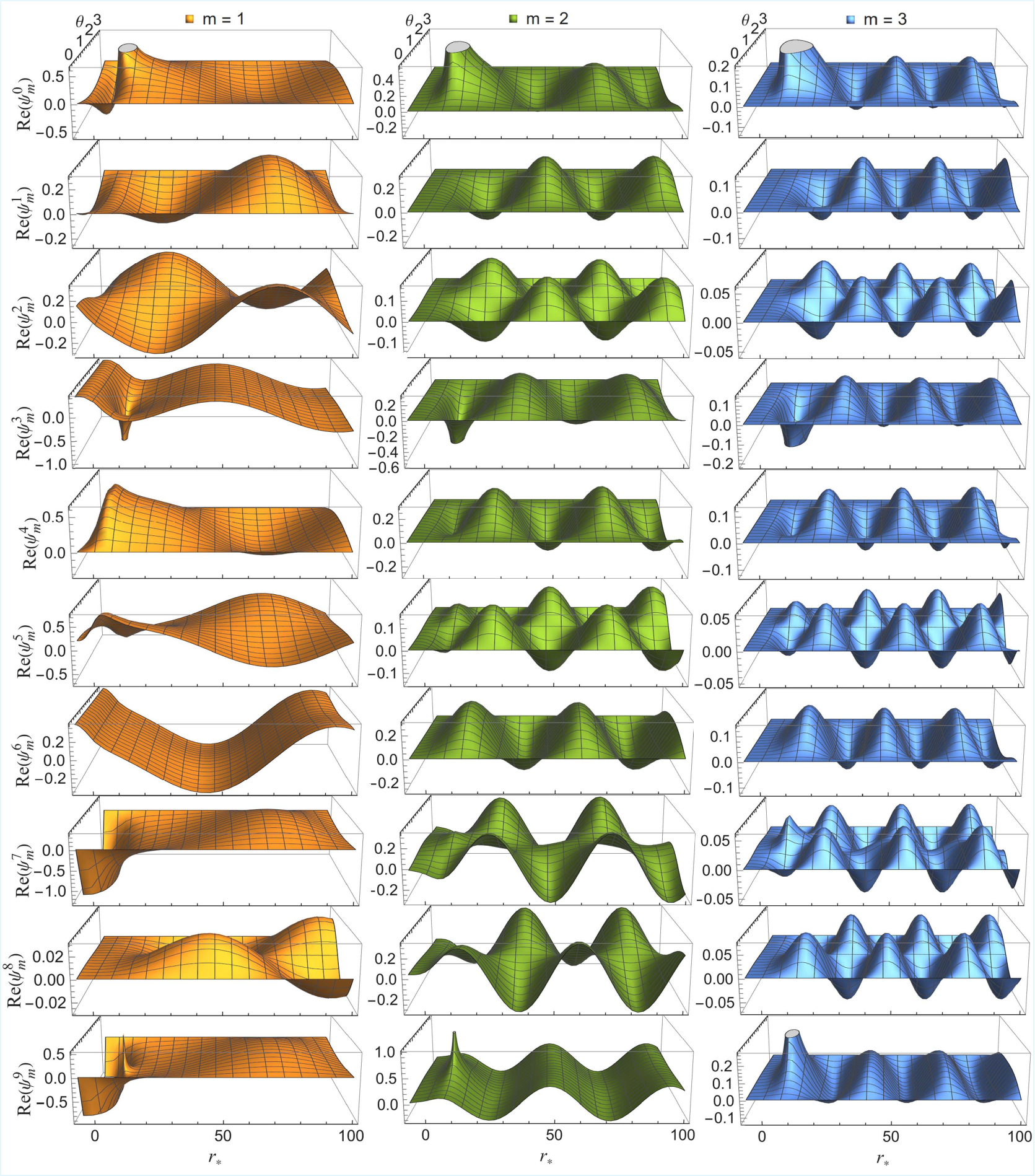}
\caption{\label{fig:3Dplots} Final $r_0 = 6M$ numerical results for the $m=1$, $2$, and $3$ modes of the retarded metric perturbation as functions of $r_*$ and $\theta$ are shown. The 3D nature of these plots is valuable to show which $\vec{\psi}_m$ components satisfy Neumann vs. Dirichlet boudary conditions at $\theta=0$ and $\theta=\pi$ consistent with Eqs.~\eqref{eq:thA}-\eqref{eq:thB}. Notice that the $\psi^{0,3,9}_m$ components of the retarded field, which are proportional to $\bar{h}^m_{tt}$, $\bar{h}^m_{t\phi}$, and $\bar{h}^m_{\phi\phi}$ respectively, are singular approaching the particle. The scale factors present in Eq.~\eqref{eq:vec} to convert from $\bar{h}^m_{\mu\nu}$ to $\vec{\psi}_m$ cause the numerical values of the retarded fields to approach constant amplitudes near the radial boundaries. Although these plots have a scale that is truncated at $r_*=100M$, the numerical domain extends to $r_*=500M$ for accuracy.}
\end{figure*}

\begin{figure}
\includegraphics[width=3.4in]{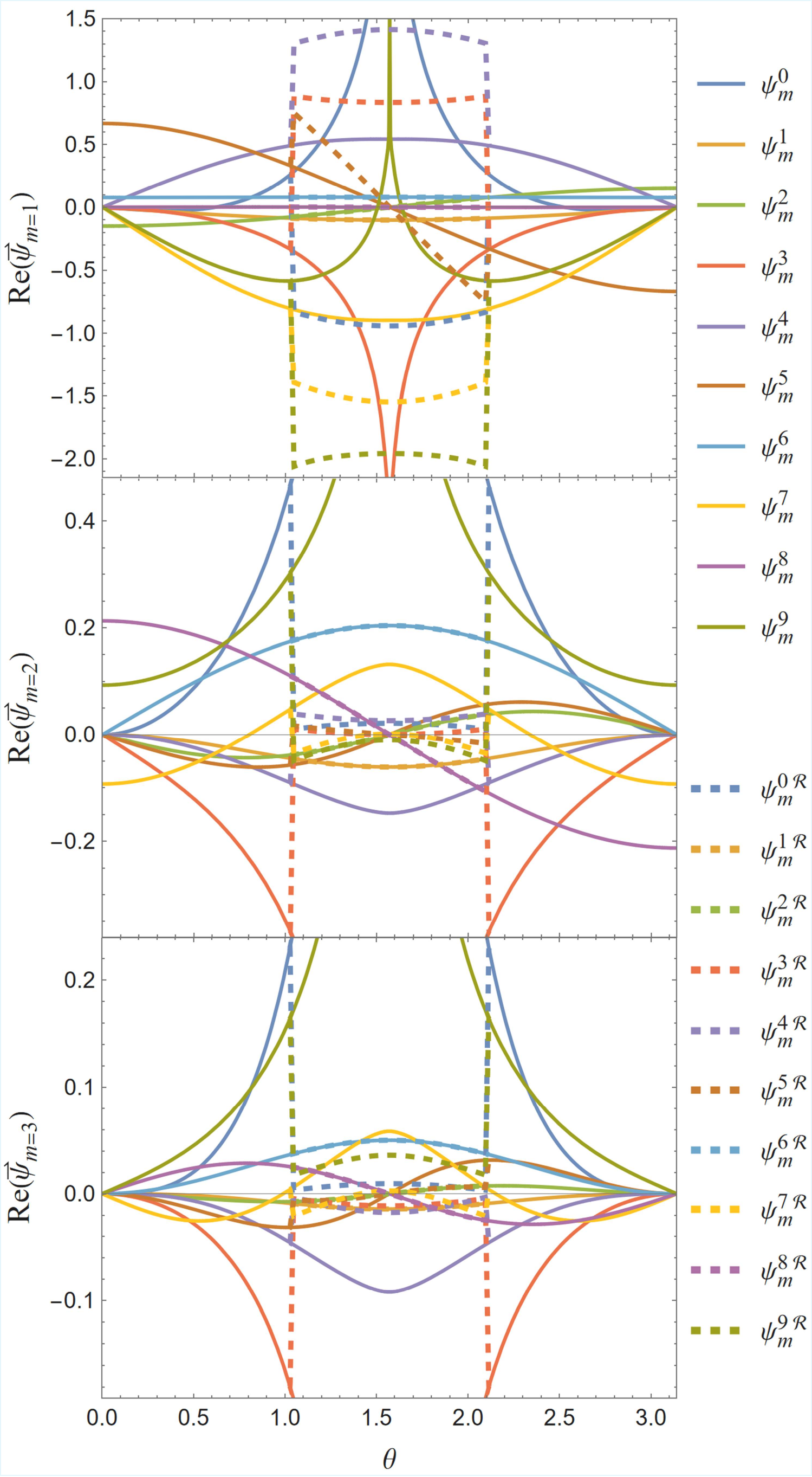}
\caption{\label{fig:2Dplots} Final $r_0 = 6M$ numerical results for the $m=1$, $2$, and $3$ modes of the retarded metric perturbation with $r=r_0$ cross-sections as functions of $\theta$ are shown. Both the retarded (solid curves) and residual (dashed curves) fields are shown. Notice that the $\psi^{2,5,8}_m$ components, which are proportional to $\bar{h}^m_{t\theta}$, $\bar{h}^m_{r\theta}$, and $\bar{h}^m_{\theta\phi}$ respectively, are anti-symmetric upon reflection across the equatorial plane, while the other metric components are symmetric. }
\end{figure}

Figure~\ref{fig:3Dplots} shows our final $r_0 = 6M$ numerical results for the $m=1$, $2$, and $3$ modes of the retarded metric perturbation as functions of $r_*$ and $\theta$. Figure~\ref{fig:2Dplots} shows $r=r_0$ cross-sections of those same results as functions of $\theta$, but with the inclusion of the residual fields. In Fig.~\ref{fig:3Dplots} we transform to the retarded solution inside the worldtube by adding the puncture values to the residual field during post-processing. Both Figure~\ref{fig:3Dplots} and Figure~\ref{fig:2Dplots} are valuable to show which $\vec{\psi}_m$ components satisfy Neumann vs. Dirichlet boundary conditions at $\theta=0$ and $\theta=\pi$ consistent with Eqs.~\eqref{eq:thA}-\eqref{eq:thB}. Notice that the $\psi^{0,3,9}_m$ components of the retarded field, which are proportional to $\bar{h}^m_{tt}$, $\bar{h}^m_{t\phi}$, and $\bar{h}^m_{\phi\phi}$ respectively, are singular approaching the particle. The scale factors present in Eq.~\eqref{eq:vec} to convert from $\bar{h}^m_{\mu\nu}$ to $\vec{\psi}_m$ cause the numerical values of the retarded fields to approach constant amplitudes near the radial boundaries. Figure~\ref{fig:2Dplots} clearly illustrates how the $\psi^{2,5,8}_m$ components, which are proportional to $\bar{h}^m_{t\theta}$, $\bar{h}^m_{r\theta}$, and $\bar{h}^m_{\theta\phi}$ respectively, are anti-symmetric upon reflection across the equatorial plane, while the other metric components are symmetric (which is consistent with expected behavior).

\subsection{Fluxes and convergence}
\label{sec:flux}

The numerical feature of our solution that we will use to demonstrate convergence is the outgoing energy flux, $\langle \dot{E}^+ \rangle$. This is partly motivated by future second-order self-force applications, where the outgoing energy flux is the basis of second-order dissipative self-force calculations (and so showing that these methods successfully determine $\langle \dot{E}^+ \rangle$ demonstrates they are likely viable for second-order self-force calculations). 

\begin{figure}
\includegraphics[width=3.4in]{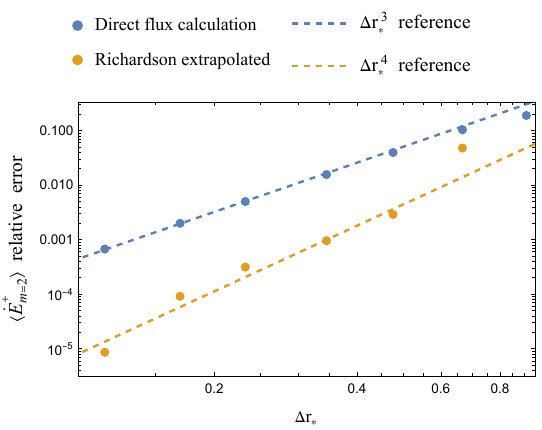}
\caption{\label{fig:converge} Convergence of the energy flux with increasing resolution is shown for the $r_0=6M$ and $m=2$ case. Richardson extrapolation enhances the convergence rate. We believe the $\theta$ integral to calculate the flux causes the $\mathcal{O}(\Delta r_*^2)$ error contribution to cancel. }
\end{figure}

The expression for $\langle \dot{E}^+ \rangle$ follows from considering how the first-order metric perturbation influences the Einstein tensor, $G_{\mu\nu}$: the second-order contribution to $G_{\mu\nu}$ from the first-order perturbation behaves as an effective stress-energy tensor that can be averaged to determine fluxes (often the Isaacson effective stress-energy tensor is used~\cite{Isaacson_1968}). Then $\langle \dot{E}^+ \rangle$ is calculated by integrating $G_{tr}$ over a large radius sphere and averaging over one period
\begin{align}
&\langle \dot{E}^+ \rangle = \sum_{m=1}^{\infty} 2 \, \langle \dot{E}^+_m \rangle \, ,
\\
&\langle \dot{E}^+_m \rangle = \frac{m^2\Omega^2}{32}\int_0^\pi \Big( |\psi^7_m|^2+4\, |\psi^8_m|^2+|\psi^9_m|^2
\\& \qquad\qquad\qquad\qquad\;\;\;\; - \psi^{7}_m \psi^{9*}_m - \psi^{9}_m \psi^{7*}_m \Big)\Big|_{r\rightarrow \infty} \sin{\theta} \, d\theta \, . \notag
\end{align}
The $m=2$ mode requires the most numerical accuracy because it has the largest flux contribution. One way to improve accuracy is to fit for the asymptotic coefficient of $\vec{\psi}_m$ based on its value and $r_*$ derivative at $r_*^\text{max}$ to extend the flux calculation deeper towards $r \simeq \infty$. Another way to enhance accuracy is to perform Richardson extrapolation as the resolution is increased; see Figure~\ref{fig:converge} to view convergence of $\langle \dot{E}^+_{m=2} \rangle$ after Richardson extrapolation to approximately 4 significant figures. We apply these two techniques to optimize accuracy for the four largest flux contributions, $m=2,3,4,5$, but the other $m$-modes have small enough fluxes that it is sufficient for us to use their raw values. Table~\ref{tab:fluxes} shows numerical comparisons for the total flux at various orbital radii. 

\begin{table}
\caption{\label{tab:fluxes}
Total fluxes for various orbital radii compared to reliable results from the Black Hole Perturbation Toolkit~\cite{BHPToolkit} Teukolsky package~\cite{1wardell_2025_15745889}. We estimate significant digits based on the convergence with increasing resolution and convergence with an increasing number of $m$-modes. Only significant digits are displayed and we have truncated the comparison numbers at 10 digits. Note that we are able to achieve almost four significant digits of flux accuracy, which should be sufficient for second-order self-force applications.
}
\begin{ruledtabular}
\begin{tabular}{c|c|c}
 $r_0 /M$ & $\langle \dot{E}^+ \rangle M^2/\mu^2 \;\; \text{(this work)} \;\;\;\;$ & BHPToolkit~\cite{BHPToolkit} $\;$ $\displaystyle \vphantom{\frac{a}{b}}$ \\ 
 \hline
 6 & $9.374\times 10^{-4}$ & $9.372704106\times 10^{-4}$ \\
 8 & $1.960\times 10^{-4}$ & $1.959794791\times 10^{-4}$ \\
 10 & $6.151\times 10^{-5}$ & $6.150372549\times 10^{-5}$
\end{tabular}
\end{ruledtabular}
\end{table}

\section{Discussion and Future Directions}

We have calculated Schwarzschild metric perturbations via $m$-modes in the frequency domain for the first time. Our approach is based on solving a system of 10 coupled elliptic PDEs in the Lorenz gauge, which we believe is a promising strategy for future second-order self-force calculations. More specifically, all the ingredients in our model handle the more general and exciting case of Kerr metric perturbations, but our preliminary Kerr results exhibited incorrect behavior near the horizon so we focused on the Schwarzschild case as a laboratory to investigate improved horizon boundary conditions. In our \texttt{Mathematica} code implementation, we found and implemented strategies to obtain the correct near-horizon behavior and are able to calculate the outgoing energy flux to approximately 4 significant digits (which is more than sufficient for future second-order calculations). There are some limitations of these \texttt{Mathematica} results including skipping the non-radiative $m=0$ mode and rougher numerical methods that have limited our overall accuracy and efficiency. Although these results are enabling valuable downstream applications, it could be said that this work partly serves as a progress report to investigate issues that we have not yet overcome entirely.

In some of our previous scalar field work~\cite{Osburn_2022}, we optimistically said ``we believe the same approach will successfully access the Lorenz gauge gravitational self-force on a compact mass orbiting a Kerr black hole" after noting that prior work involving the Lorenz gauge PDEs in the time-domain~\cite{Dolan_2013} (which inspired this frequency domain approach) encountered sufficiently large obstacles to postpone their full Kerr implementation; it seems as if the Lorenz gauge PDEs are resistant to be solved without a fight because a different set of obstacles has similarly delayed our full Kerr results. However, there are still reasons for optimism. Although it was not presented here, we have been simultaneously developing a version of the Lorenz gauge Kerr perturbation PDE solver based on \texttt{SpECTRE} initial data routines~\cite{Vu_2026}. Our preliminary \texttt{SpECTRE} results suggest that spectral methods and compactification overcome the near-horizon issues described here. We are working towards a full open-source \texttt{SpECTRE} implementation that we will use to generate Kerr metric perturbation results with improved accuracy and efficiency.

Besides the full Kerr \texttt{SpECTRE} implementation that is in progress, the most obvious follow-up project will be to use elliptic PDE methods to calculate the Kerr second-order self-force for the first time; there is an ongoing effort to calculate the $m$-modes of the Kerr second-order source, and with only minor modifications this approach should be able to accept such a second-order source as an input and then output second-order metric perturbations that will enable new and improved second-order waveform models. Other potential future directions include using hyperboloidal slicing and compactification, considering the case of eccentric and/or inclined orbits, and further improving the realism by introducing features such as environmental effects and/or a spinning secondary.

\begin{acknowledgments}
We thank Nami Nishimura, Savitri Iyer, Nils Vu, Jonathan Thompson, and Benjamin Leather for helpful discussions, as well as Sam Dolan and Leor Barack for sharing time-domain $m$-mode data and helpful discussions.
TO gratefully acknowledges support from the National Science Foundation under Grant No.\ PHY-2309020 at SUNY Geneseo and from the Fulbright U.S. Scholar Program, which is sponsored by the U.S. Department of State and the Fulbright Commission in Ireland; the contents are solely the responsibility of the authors and do not necessarily represent the official views of the Fulbright Program, the U.S. Government, or the Fulbright Commission in Ireland.
This work is supported by ERC grant EMRIWaveforms (DOI: \href{https://doi.org/10.3030/101200625}{10.3030/101200625}). This work makes use of the Black Hole Perturbation Toolkit.

\end{acknowledgments}

\appendix

\onecolumngrid

\section{Matrices for system of PDEs}
\label{sec:pdes}

The details associated with the form of the field equations used in our numerical calculations are presented here. Recall Eq.~\eqref{eq:PDEs} (reproduced here):
\begin{align}
\left(\frac{\partial^2}{\partial r_*^2} + \frac{\Delta}{r^4} \frac{\partial^2}{\partial \theta^2}  + \mathbf{A}_m \frac{\partial}{\partial r_*}  + \mathbf{B}_m \frac{\partial}{\partial \theta}  + \mathbf{C}_m \right) \vec{\psi}_m = \vec{S}_m \, .
\end{align}
The $\mathbf{A}_m$, $\mathbf{B}_m$, and $\mathbf{C}_m$ matrices follow from substituting the above into Eqs.~\eqref{eq:vec}, \eqref{eq:modes}, and \eqref{eq:fieldeqs}:
\begin{align}
& \mathbf{A}_m = -\frac{2}{r^2} \left(\begin{array}{cccccccccc} 
2 & 0 & 0 & 0 & 0 & 0 & 0 & 0 & 0 & 0 \\ 
0 & 2 & 0 & 0 & 0 & 0 & 0 & 0 & 0 & 0 \\ 
0 & 0 & 1 & 0 & 0 & 0 & 0 & 0 & 0 & 0 \\ 
0 & 0 & 0 & 1 & 0 & 0 & 0 & 0 & 0 & 0 \\ 
0 & 0 & 0 & 0 & 2 & 0 & 0 & 0 & 0 & 0 \\ 
0 & 0 & 0 & 0 & 0 & 1 & 0 & 0 & 0 & 0 \\
0 & 0 & 0 & 0 & 0 & 0 & 1 & 0 & 0 & 0 \\ 
0 & 0 & 0 & 0 & 0 & 0 & 0 & 0 & 0 & 0 \\ 
0 & 0 & 0 & 0 & 0 & 0 & 0 & 0 & 0 & 0 \\ 
0 & 0 & 0 & 0 & 0 & 0 & 0 & 0 & 0 & 0 
\end{array}\right),
\\
& \mathbf{B}_m =  \frac{\Delta}{r^4\tan{\theta}}\mathbf{I} - \frac{2\Delta}{r^4} \left(\begin{array}{cccccccccc} 
0 & 0 & 0 & 0 & 0 & 0 & 0 & 0 & 0 & 0 \\ 
0 & 0 & \frac{\Delta}{r^2} & 0 & 0 & 0 & 0 & 0 & 0 & 0 \\ 
0 & -1 & 0 & 0 & 0 & 0 & 0 & 0 & 0 & 0 \\ 
0 & 0 & 0 & 0 & 0 & 0 & 0 & 0 & 0 & 0 \\ 
0 & 0 & 0 & 0 & 0 & \frac{2\Delta}{r^2} & 0 & 0 & 0 & 0 \\ 
0 & 0 & 0 & 0 & -1 & 0 & 0 & \frac{\Delta}{r^2} & 0 & 0 \\ 
0 & 0 & 0 & 0 & 0 & 0 & 0 & 0 & \frac{\Delta}{r^2} & 0 \\
0 & 0 & 0 & 0 & 0 & -2 & 0 & 0 & 0 & 0 \\ 
0 & 0 & 0 & 0 & 0 & 0 & -1 & 0 & 0 & 0 \\ 
0 & 0 & 0 & 0 & 0 & 0 & 0 & 0 & 0 & 0 
\end{array}\right),
\\
& \mathbf{C}_m = \left(\omega^2 - \frac{m^2\Delta}{r^4\text{sin}^2\theta} \right)\mathbf{I} - \frac{2\Delta}{r^4} \times
\\& \left(
\begin{array}{cccccccccc}
 \frac{1-r}{\Delta } & \frac{2 i r^2 \omega }{\Delta } & 0 & 0 & \frac{3\Delta+r^2}{2 r\Delta} & 0 & 0 & \frac{-\Delta }{r^3} & 0 & \frac{-\Delta }{r^3} \vphantom{\frac{H^2}{j_2}}\\
 \frac{i r^2 \omega }{\Delta } & \frac{r-1}{r} & \frac{\Delta  \cos {\theta}}{r^2 \sin{\theta}} & \frac{im \Delta}{r^2 \sin{\theta}} &
   \frac{i r^2 \omega }{\Delta } & 0 & 0 & 0 & 0 & 0 \vphantom{\frac{H^2}{j_2}}\\
 0 & 0 & \frac{\Delta +2 r \cos{2 \theta}}{2 r^2\, \text{sin}^2\theta} & \frac{i m \cos{\theta}}{\text{sin}^2\theta} & 0 & \frac{i r^2 \omega
   }{\Delta } & 0 & 0 & 0 & 0  \vphantom{\frac{H^2}{j_2}}\\
 0 & \frac{-i m}{\sin{\theta}} & \frac{-i m \cos {\theta}}{\text{sin}^2\theta} & \frac{\Delta +2 r \cos{2 \theta}}{2 r^2\, \text{sin}^2\theta} &
   0 & 0 & \frac{i r^2 \omega }{\Delta } & 0 & 0 & 0  \vphantom{\frac{H^2}{j_2}}\\
 \frac{3\Delta+r^2}{2 r\Delta} & \frac{2 i r^2 \omega }{\Delta } & 0 & 0 & \frac{(r-3) (2 r-3)}{\Delta } & \frac{2 \Delta \cos{\theta}}{r^2 \sin{\theta}} &
   \frac{2 im \Delta}{r^2 \sin{\theta}} & \frac{\Delta (3-r)}{r^3} & 0 & \frac{\Delta (3-r)}{r^3}  \vphantom{\frac{H^2}{j_2}}\\
 0 & 0 & \frac{i r^2 \omega }{\Delta } & 0 & 0 & \frac{1}{2\, \text{sin}^2\theta}\scriptstyle{-}\textstyle{\frac{6}{r}}\scriptstyle{+2} & \frac{i m \cos{\theta}}{\text{sin}^2\theta} &
   \frac{\Delta  \cos{\theta}}{r^2 \sin{\theta}} & \frac{im \Delta}{r^2 \sin{\theta}} & \frac{-\Delta  \cos{\theta}}{r^2 \sin{\theta}}  \vphantom{\frac{H^2}{j_2}}\\
 0 & 0 & 0 & \frac{i r^2 \omega }{\Delta } & \frac{-i m}{\sin{\theta}} & \frac{-i m \cos{\theta}}{\text{sin}^2\theta} & \frac{1}{2 \, \text{sin}^2\theta}\scriptstyle{-}\textstyle{\frac{6}{r}}\scriptstyle{+2} & 0 & \frac{2 \Delta  \cos{\theta}}{r^2 \sin{\theta}} & \frac{im \Delta}{r^2 \sin{\theta}}  \vphantom{\frac{H^2}{j_2}}\\
 \frac{-r}{\Delta } & 0 & 0 & 0 & \frac{(3-r) r}{\Delta } & 0 & 0 & \frac{\cos{2 \theta}+2 r-1}{2 r\, \text{sin}^2\theta} & \frac{2 i m \cos{\theta}}{\text{sin}^2\theta} & \frac{\Delta }{r^2}\scriptstyle{-}\textstyle{\frac{1}{\text{sin}^2\theta}}  \vphantom{\frac{H^2}{j_2}}\\
 0 & 0 & 0 & 0 & 0 & \frac{-i m}{\sin{\theta}} & \frac{\cos{\theta}}{\sin{\theta}} & \frac{-i m \cos{\theta}}{\text{sin}^2\theta} &
   \frac{2}{\text{sin}^2\theta}\scriptstyle{+}\textstyle{\frac{1}{r}}\scriptstyle{-1} & \frac{i m \cos{\theta}}{\text{sin}^2\theta}  \vphantom{\frac{H^2}{j_2}}\\
 \frac{-r}{\Delta } & 0 & 0 & 0 & \frac{(3-r) r}{\Delta } & \frac{-2 \cos{\theta}}{\sin{\theta}} & \frac{-2 i m}{\sin{\theta}} &
   \frac{\Delta }{r^2}\scriptstyle{-}\textstyle{\frac{1}{\text{sin}^2\theta}} & \frac{-2 i m \cos{\theta}}{\text{sin}^2\theta} & \frac{\cos{2 \theta}+2 r-1}{2 r\, \text{sin}^2\theta} \vphantom{\frac{H^2}{j_2}}
\end{array}
\right), \notag
\end{align}
where we have set $M=1$ and introduced $\omega = m\Omega$. Some of these have a relatively simple form, but all three matrices are much more complicated for the Kerr case. One example is $\mathbf{A}_m$, which is diagonal for Schwarzschild but not for Kerr. Our \texttt{Mathematica} code implements these matrices as functions that accept a single $\theta$ input and a vector of many $r$ inputs to return a 3D array containing all the matrix elements for every $r$ but only the single $\theta$. Repeating that process for each $\theta$ generates a set of 3D arrays that are reshaped to provide the coefficients for the linear system of finite difference equations.

\section{Eigenvector analysis of near-horizon behavior}
\label{sec:eigenvecs}

Recall that the field equations simplify near the horizon according to Eq.~\eqref{eq:horeqs} (reproduced here)
\begin{align}
\left(\frac{\partial^2}{\partial r_*^2}   + \mathbf{A}_m \big|_{r=2M} \frac{\partial}{\partial r_*}  + \mathbf{C}_m \big|_{r=2M} \right) \vec{\psi}_m = 0 \, ,
\end{align}
which is a system of coupled ODEs that can be written in first-order form according to Eq.~\eqref{eq:ODEs} (reproduced here)
\begin{align}
\frac{\partial}{\partial r_*} \left(\begin{array}{c} 
\vec{\psi}_m \\\,\\ \frac{\partial}{\partial r_*} \vec{\psi}_m
\end{array}\right) = \left(\begin{array}{cc}
\mathbf{0} & \mathbf{I} \\\,&\,\\ -\mathbf{C}_m & -\mathbf{A}_m
\end{array}\right)\left(\begin{array}{c} 
\vec{\psi}_m \\\,\\ \frac{\partial}{\partial r_*} \vec{\psi}_m
\end{array}\right) \, ,
\end{align}
and there are 20 independent homogeneous solutions associated with the eigenvalues and eigenvectors of the $20\times 20$ matrix according to Eq.~\eqref{eq:homo} (reproduced here)
\begin{align}
\left(\begin{array}{c} 
\vec{\psi}_m \\\,\\ \frac{\partial}{\partial r_*} \vec{\psi}_m
\end{array}\right) = \left(\begin{array}{c} 
\vec{v} \\\,\\ \lambda \, \vec{v}
\end{array}\right) \, e^{\lambda r_*}\, ,
\end{align}
where $\lambda$ represents one of the 20 eigenvalues that describe the $r_*$ exponent (which encodes radial dependence near the horizon) and $\vec{v}$ represents an associated eigenvector. Table~\ref{tab:eigenvals} presented the eigenvalues, but here we also present the eigenvectors in Table~\ref{tab:eigenvectors}. Based on how the eigenvector elements are related to metric perturbation components, it can be shown that only solutions with $Re(\lambda)=0$ eigenvalues satisfy the Lorenz gauge condition.

\begin{table}
\caption{\label{tab:eigenvectors}
The 20 eigenvalues ($\lambda$) and eigenvectors ($\vec{v}$) representing the near-horizon homogeneous solutions of the field equations are shown. Technically, the eigenvectors of the $20\times 20$ matrix will each have $20$ elements with $\vec{v}$ representing the first 10 elements and $\lambda\,\vec{v}$ representing the last 10 elements, but we only present $\vec{v}$ for brevity. Based on how the eigenvector elements are related to metric perturbation components, it can be shown that only solutions with $Re(\lambda)=0$ eigenvalues satisfy the Lorenz gauge condition.
}
\begin{ruledtabular}
\begin{tabular}{c|cccccccccccccccccccc}
$Im(\lambda)$ & $-i\omega$ & $-i\omega$ & $-i\omega$ & $-i\omega$ & $-i\omega$ & $-i\omega$ & $-i\omega$ & $-i\omega$ & $-i\omega$ & $-i\omega$ & $+i\omega$ & $+i\omega$ & $+i\omega$ & $+i\omega$ & $+i\omega$ & $+i\omega$ & $+i\omega$ & $+i\omega$ & $+i\omega$ & $+i\omega$ \\
$Re(\lambda)$ & 0 & 0 & 0 & 0 & 0 & 0 & $\frac{1}{2}$ & $\frac{1}{2}$ & $\frac{1}{2}$ & 1 & 0 & 0 & 0 & 0 & 0 & 0 & $\frac{1}{2}$ & $\frac{1}{2}$ & $\frac{1}{2}$ & 1$\vphantom{\frac{H^2}{j_2}}$\\
 \hline
$v_0$ & 1 & 0 & 0 & 0 & 0 & 0 & 0 & 0 & $\scriptstyle{i\omega-}\textstyle{\frac{1}{4}}$ & 1 & 1 & 0 & 0 & 0 & 0 & 0 & 0 & 0 & $\scriptstyle{i\omega+}\textstyle{\frac{1}{4}}$ & 1 \\ 
$v_1$ & 1 & 0 & 0 & 0 & 0 & 0 & 0 & 0 & 0 & -1 & -1 & 0 & 0 & 0 & 0 & 0 & 0 & 0 & 0 & 1 \\ 
$v_2$ & 0 & 1 & 0 & 0 & 0 & 0 & 1 & 0 & 0 & 0 & 0 & 1 & 0 & 0 & 0 & 0 & 1 & 0 & 0 & 0 \\ 
$v_3$ & 0 & 0 & 1 & 0 & 0 & 0 & 0 & 1 & 0 & 0 & 0 & 0 & 1 & 0 & 0 & 0 & 0 & 1 & 0 & 0 \\ 
$v_4$ & 1 & 0 & 0 & 0 & 0 & 0 & 0 & 0 & $\scriptstyle{-i\omega+}\textstyle{\frac{1}{4}}$ & 1 & 1 & 0 & 0 & 0 & 0 & 0 & 0 & 0 & $\scriptstyle{-i\omega-}\textstyle{\frac{1}{4}}$ & 1 \\ 
$v_5$ & 0 & 1 & 0 & 0 & 0 & 0 & -1 & 0 & 0 & 0 & 0 & -1 & 0 & 0 & 0 & 0 & 1 & 0 & 0 & 0 \\
$v_6$ & 0 & 0 & 1 & 0 & 0 & 0 & 0 & -1 & 0 & 0 & 0 & 0 & -1 & 0 & 0 & 0 & 0 & 1 & 0 & 0 \\ 
$v_7$ & 0 & 0 & 0 & 1 & 0 & 0 & 0 & 0 & $\frac{1}{2}$ & 0 & 0 & 0 & 0 & 1 & 0 & 0 & 0 & 0 & $\scriptstyle{-}\textstyle{\frac{1}{2}}$ & 0 \\ 
$v_8$ & 0 & 0 & 0 & 0 & 1 & 0 & 0 & 0 & 0 & 0 & 0 & 0 & 0 & 0 & 1 & 0 & 0 & 0 & 0 & 0 \\ 
$v_9$ & 0 & 0 & 0 & 0 & 0 & 1 & 0 & 0 & $\frac{1}{2}$ & 0 & 0 & 0 & 0 & 0 & 0 & 1 & 0 & 0 & $\scriptsize{-}\textstyle{\frac{1}{2}}$ & 0 
\end{tabular}
\end{ruledtabular}
\end{table}

\bibliography{main}

\end{document}